\documentclass[12pt,fleqn, a4paper]{article}

\newcommand{\ra}{\rightarrow}

\newcommand{\bra}{\langle} \newcommand{\ket}{\rangle}
\newcommand{\be}{\begin{equation}}
\newcommand{\ee}{\end{equation}}
\newcommand{\bea}{\begin{eqnarray}}
\newcommand{\eea}{\end{eqnarray}}

\newcommand{\E}{\mbox{e}}

\newcommand{\ffi}{\varphi}

\newcommand{\ep}{\qquad {\vrule height 10pt width 8pt depth 0pt}}

\newcommand{\grintl}{[\kern-.18em [}
\newcommand{\grintr}{]\kern-.18em ]}

\newtheorem{lem}{Lemma}[section]
\newtheorem{prop}{Proposition}[section]
\newtheorem{thm}{Theorem}[section]
\newtheorem{cor}{Corollary}[section]
\usepackage{amsfonts}
\def\R{{\mathbb R}}

\def\0{{\mathbb O}}
\def\un{{\mathbb I}}
\def\T{{\mathbb T}}
\def\Z{{\mathbb Z}}
\def\N{{\mathbb N}}
\def\C{{\mathbb C}}
\def\E{{\mathbb E}}
\def\P{{\mathbb P}}

\begin{document}

\title{Fractional Moment Estimates for Random Unitary Operators}
\author{Alain Joye\\ \\
Institut Fourier
\\ Universit\'e de Grenoble 1, BP 74,
\\38402 Saint-Martin d'H\`eres Cedex, France}
\date{  }
\maketitle
\abstract{We consider unitary analogs of $d-$dimensional Anderson
models on $l^2(\Z^d)$ defined by the product $U_\omega=D_\omega S$
where $S$ is a deterministic unitary  and $D_\omega$ is a diagonal matrix 
of i.i.d. random phases. The operator $S$ is an absolutely continuous 
band matrix which depends on parameters controlling the size of its 
off-diagonal elements. 
We adapt the method of Aizenman-Molchanov to get exponential
estimates on fractional moments of the matrix elements 
of $U_\omega(U_\omega -z)^{-1}$, provided the distribution of phases is 
absolutely continuous and the parameters correspond
to small off-diagonal elements of $S$.
Such estimates imply almost sure localization 
for $U_\omega$.}


\setcounter{equation}{0}
\section{Introduction}

Unitary operators displaying a band structure with respect
to a distinguished basis appear in the description of the
long time  properties of certain quantum dynamical 
systems. For example,
such operators on $l^2(\N)$ are used to model the dynamics of an
electron in a ring threaded by a time dependent magnetic flux.
In some regime of the physical parameters, 
certain phases of the matrix elements can be considered as 
random variables. These models are useful for numerical 
investigations. See \cite{bb}, \cite{bhj} and references 
therein for details on the model and more on
quantum dynamical systems.

Unitary operators with a similar band structure appear naturally
in the study of orthogonal polynomials on the unit circle $S^1$
with respect to a measure $d\mu$ on the torus $\T$, see \cite{s}. 
Indeed, it is shown in \cite{cmv} that multiplication by 
$e^{i\alpha}\in S^1$ on 
$L^2(\T, d\mu(\alpha))$ expressed in a
certain basis of orthonormal polynomials is represented by such a band
matrix in $l^2(\N)$. This construction is simpler than the earlier 
Hessenberg form of the matrix 
representation of this unitary operator provided in \cite{gt}.
The spectral analysis of the unitary operator 
therefore yields informations on the polynomials. Considering some 
phases as random amounts to considering certain types of 
random polynomials.

The spectral analysis of a certain set of deterministic and 
random unitary operators with a band structure is undertaken 
in \cite{bhj} and \cite{j}. This set contains 
the examples mentionned above as particular cases. 
In the random cases studied in these two papers, the operators 
considered consist in matrices on $l^2(\Z)$ (which are unitarily equivalent 
to matrices) of the following form: 
$U_\omega=D_\omega S$ where $S$ is a deterministic unitary 
and $D_\omega$ is a diagonal matrix of random phases, see \cite{j}. 
The operator $S$ is an absolutely continuous 
band matrix which depends on a parameter
$t\in ]0,1[$ which controls the size of its off-diagonal 
elements, see Section \ref{mod}.
When the phases are i.i.d random variables, typical results 
obtained for discrete one-dimensional random Schr\"odinger operators
are shown in \cite{bhj} and \cite{j} to hold in the unitary 
setting as well. For instance, the availability of a transfer 
matrix formalism to express generalized eigenvectors 
allows to introduce a Lyapunov exponent, to prove a unitary version 
of Ishii-Pastur Theorem, and get absence of absolutely continuous 
spectrum \cite{bhj}. A density of states can be introduced and
a Thouless formula is proven in \cite{j}.
Related analyses in the framework of orthogonal polynomials on
the unit circle are provided in \cite{gt}, \cite{t}, \cite{s}.  

In the present paper, we introduce a natural generalization of such unitary
operators to higher dimensions, i.e. to $l^2(\Z^d)$, $d\geq 1$, 
in analogy with the self-adjoint Anderson model. The construction is
motivated by the structure of $U_\omega$ given as a product of a diagonal random
operator $D_\omega$ times a deterministic unitary $S$.
This structure is a natural transposition to the 
unitary setting of that of the Anderson model consisting 
in the sum of a diagonal random potential and the deterministic discrete
Laplacian. The extension is straightforward and consists in matrices 
$U_\omega$ of the form $D_\omega S$, acting on $l^2(\Z^d)$, 
where the infinite matrices $D_\omega$ and $S$ have similar properties
with respect to the canonical basis of $l^2(\Z^d)$, see Section \ref{mod}.
In particular, we assume the phases in the diagonal of $D_\omega$ are 
i.i.d. with an absolutely continuous distribution, and the operator $S$ depends
now on a set of $d$ parameters $(t_1, t_2, \cdots, t_d)$ which control 
the size of its off-diagonal elements. 

Once defined, these random operators call for an analysis of their spectral 
properties. In the self-adjoint case, the localization properties of the 
$d$-dimensional Anderson model can be conveniently proven for large
disorder by means of the fractional moment method of Aizenman and 
Molchanov \cite{am} and the Simon-Wolff criterion \cite{sw}. 
Our main result, Theorem \ref{efm} below,  is an exponential estimate
on the fractional moments of the matrix elements of 
$U_\omega(U_\omega - z)^{-1}$, 
uniform in $z$, obtained by an adaptation to the unitary setting of the
Aizenman-Molchanov method. Our estimate holds for a range of  parameters 
$(t_1, t_2, \dots, t_d)$ such that the off-diagonal elements of $S$ are small
enough. This last condition is the equivalent in our setting of the large
disorder assumption made in the self-adjoint case. Then we apply the unitary
version of the Simon-Wolff criterion proven by Combescure in \cite{c} to 
derive localization for $U_\omega$ in Corollary \ref{uloc}, for the same 
range of parameters.

\section{The Model and Main Result}\label{mod}\setcounter{equation}{0}

We denote by $|k\ket = | k_1, k_2, \cdots, k_d \ket $ the unit vector at site
$k\in \Z^d$, so that $\{|k\ket\}_{k\in\Z^d}$ form an orthonormal basis of 
$L^2(\Z^d)$. We introduce a probability space $(\Omega, {\cal F}, \P)$, where 
$\Omega$ is identified with $\{{\T}^{\Z^d} \}$,
$\T$ being the torus,
and $\P=\otimes_{k\in\Z^d}\P_k$, where $\P_{k}=\P_0$ 
for any $k\in\Z^d$ is a probability distributions on $\T$, 
and ${\cal F}$ the $\sigma$-algebra 
generated by the cylinders. We introduce
a set of random vectors on 
$(\Omega, {\cal F}, \P)$  by 
\bea\label{beta}
\theta_k: \Omega \rightarrow \T,
  \ \ \mbox{s.t.} \ \ \theta_k(\omega)=\omega_{k}, \ \ \ k\in \Z^d.
\eea
These random vectors $\{\theta_k\}_{k\in\Z^d}$ are thus i.i.d on $\T$.

In the one dimensional case, $d=1$,  we consider 
unitary operators of the form
\be\label{1d}
 U_{\omega}=D_{\omega}S_0, \,\,\,\mbox{ with } 
D_{\omega}=\mbox{ diag }\{e^{-i\theta_k(\omega)}\}
\ee
and 
\be\label{s0}
S_0=\pmatrix{\ddots & rt & -t^2& & & \cr
              & r^2& -rt  & & & \cr
              & rt & r^2 & rt & -t^2& \cr
              & -t^2 &-tr & r^2& -rt& \cr 
              & & & rt &r^2 & \cr
              & & & -t^2& -tr&\ddots },
\ee 
where the translation along the diagonal is fixed by 
$\bra \ffi_{2k-2}|S_0\ffi_{2k}\ket =-t^2$, $k\in\Z$. The parameters $t$ and $r$ are
linked by $r^2+t^2=1$ to ensure unitarity. We shall sometimes write $S_0(t)$ to 
emphasize this dependence. 
The spectrum of $S_0(t)$ is purely absolutely continuous and consists in the set
\be
\sigma(S_0(t))=\Sigma_0(t)=\{e^{\pm i \arccos (1-t^2(1+\cos(y))) }, y\in \T\}.
\ee
For this and other properties of $S_0$, relations between $U_\omega$ with 
the physical model alluded to in Section 1 or links with orthogonal 
polynomials, see \cite{j}. Note that the band structure (\ref{s0}) is the 
simplest one a unitary operator can take without being trivial from the point 
of view of its spectrum,  \cite{bhj}.\\

To deal with $d$-dimensional operators, we introduce the following 
natural generalization of (\ref{1d}) to $l^2(\Z^d)$. 
We consider the unitary 
\be \label{defu}
U_\omega = D_\omega S \ \ \ \mbox{on }\ \ l^2(\Z^d),
\ee 
where $D_\omega$ is diagonal again
\be\label{defd}
D_{\omega} |k\ket =
e^{-i\theta_k(\omega)}|k\ket
\ee
whereas the deterministic part is defined by  
\be\label{defs}
 S=S_1\otimes \cdots \otimes S_d.
\ee
That is, we view $l^2(\Z^d)$ as $\otimes_{j=1}^d l^2(\Z)$ so that $|k\ket \simeq 
| k_1\ket \otimes \cdots \otimes |k_d \ket $ and
$S_j$ acts on $|k_j\ket$ as $S_0$ in (\ref{s0}). 
We shall identify $S_j$ with 
$\un\otimes \dots \un \otimes S_j\otimes \un \cdots \otimes \un$. 
A natural symmetric choice consists in taking the same parameter $t$ for each unitary 
$S_j(t)$ appearing in the definition
of $S=S(t)$. But we can naturally consider non-symmetric cases characterized by a set 
of parameters
$t=( t_1,t_2,\cdots, t_d)\in ]0,1[^d$ to construct the unitary operator 
$S(t)=S_1(t_1) S_2(t_2)\cdots S_d(t_d)$. Note that one gets rightaway 
that $S(t)$ is purely absolutely continuous and
\be
\sigma(S(t))=\Sigma_0(t_1)\times \Sigma_0(t_2) \times \cdots \times \Sigma_0(t_d).
\ee
Moreover, with the norm $|x|=\max_{j=1,\cdots, d}|x_j|$, $x\in \R^d$, 
we have the band structure
\be\label{bs}
\bra k | S j\ket = 0 \ \ \ \mbox{ if } \ \ |j - k|>2.
\ee

\noindent {\bf Remarks:} \\
\noindent i) In this definition, $S$ plays the role of the free $d$-dimensional Laplacian 
in the self-adjoint case. Therefore, in the same way  the Laplacian can be 
written as a sum of commuting one-dimensional Laplacians, $S$ is defined 
as a product of commuting unitary operators. \\
ii) Our construction of $S$ gives a band structure to $U_\omega$. However, 
our results do not require such a structure, see below.\\
iii) Note that as $|t|\ra 0$, $S(t)$ tends to the identity operator.
\vskip.3cm

Our main result is an estimate on the  fractional moments 
of, essentially, the matrix elements of the resolvent of $U_\omega$. 
\begin{thm}\label{efm} 
Let $U_\omega$ be defined by (\ref{defu}, \ref{defd}, \ref{defs}). Assume
that $\{\theta_k(\omega)\}_{k\in\Z^d}$ are i.i.d. and distributed according to 
the probability measure $d\nu(\theta)=\tau(\theta)d\theta$, where $\tau\in L^\infty(\T)$.
Let $s\in ]0,1[$. There exists $t_0(s)>0$ small enough and $0<K(s)<\infty$ such that 
if $|t|<t_0(s)$, there exists $\gamma(s,t)>0$ 
so that for any $j,k\in\Z^d$ and for any $z\in\C$, 
\be\label{esres}
\E(|\bra j |U_\omega (U_\omega -z)^{-1} k\ket|^s)\leq K(s)e^{-\gamma(s,t)|j-k|}.
\ee
\end{thm}
{\bf Remarks:} \\
i) The Theorem is true for more general deterministic unitary operators
$S$ than (\ref{defs}). The only requirement is that for some $\gamma(s)>0$,
\be
\sup_k \sum_{j\neq k} |\bra S k|j\ket|^se^{\gamma |k-j|} < C_\nu^{(1)}(s) \inf_k  
|\bra S k|k\ket|^s,
\ee
where $C_\nu^{(1)}(s)$ is defined in (\ref{1}) and depends on $s$ and on $\nu$ only. 
This condition corresponds to the large disorder assumption in the self-adjoint case. \\
ii) The random variables $\theta_k(\omega)$ need not be independent, and 
their distribution can be more general, see \cite{am}. However, we stick to 
the present hypotheses for simplicity.
\vskip.3cm

\begin{cor}\label{uloc} Consider $U_\omega=D_\omega S(t)$ under 
the hypotheses of Theorem \ref{efm}. Then, if $|t|< t_0(s)$,  
$$\sigma (U_\omega)\ \  \mbox{is pure point almost surely.}$$
\end{cor}
{\bf Note:}\\
As this paper was being completed, the preprint \cite{ps} appeared.
It announces that estimates of the type (\ref{esres}) are proven 
by Stoiciu in the realm of orthogonal polynomials on the unit circle 
and proves that dynamical localization is a consequence of these
estimates in this set up.\\

The rest of the paper is organized as follows. 
The next Section describes the effect of 
changing a phase at one site in terms of rank one perturbations 
in order to derive formulas for later use. Then we prove Theorem
\ref{efm} along the lines of \cite{am},  \cite{ag} in Section \ref{fm}. The 
Corollary on localization is proven  in 
Section \ref{loc}. An Appendix containing some technical material 
closes the paper.

\section{Rank One Perturbations}\setcounter{equation}{0}

By construction, the variation of a random phase at one site is described by
a rank one perturbation.  As randomness plays no particular role here, 
we drop the $\omega$'s in the notation.

Let $j\in \Z^d$ be fixed. We define $\hat{D}$ by taking $\theta_j=0$
in the definition of $D$:
\be\label{dhat}
\hat{D}=e^{i\theta_j|j\ket\bra j|}D=D +|j\ket\bra j|(1-e^{-i\theta_j})\equiv D +|j\ket\bra j|\eta_j, \ \ \mbox{with }\ \ \eta_j=1-e^{-i\theta_j},
\ee
so that, with the obvious notations,
\be\label{uhat}
\hat{U}=\hat{D}S=e^{i\theta_j|j\ket\bra j|}U =U+|j\ket\bra j|S\eta_j.
\ee
Let $z\not\in S^1$. By the first resolvent identity, we have
\bea
(\hat U -z)^{-1}-(U-z)^{-1}&=&-(\hat U -z)^{-1}|j\ket\bra j|\eta_jS(U-z)^{-1}
\nonumber\\
&=&-(U -z)^{-1}|j\ket\bra j|\eta_jS(\hat U-z)^{-1}.
\eea
Therefore, $F(z)=S(U-z)^{-1}$ and $\hat F(z)= S(\hat U -z)^{-1}$ satisfy
\be\label{r1}
\hat F(z)-F(z)=-\eta_j\hat F(z) |j\ket\bra j| F(z)=-\eta_j F(z) |j\ket\bra j|\hat F(z).
\ee
It is readily checked that this implies
\be
F(z)= \hat F(z) +\frac{\eta_j}{1-\eta_j\bra j | \hat F(z) j\ket }\hat F(z)| j\ket \bra j|\hat F(z).
\ee
Hence, with the notation $F(j,k;z)=\bra j| F(z) k\ket$ and similarly with $\hat F(z)$, for
any $j \in\Z^d$,
\be\label{b6}
F(j,k;z)=\frac{\hat F(j,k;z)}{1-\eta_j\hat F(j,j;z)}.
\ee
We emphasize that in the relation above, the operator $\hat F(z)$ 
depends on $j$ fixed.
Note also that $F(j,k;z)=e^{i\theta_j}\bra j| U(U-z)^{-1} k\ket$, so that it
is equivalent to deal with $F(z)$ or $ U(U-z)^{-1}$ as far as the modulus of
matrix elements is concerned. We choose to deal with $F(z)$ because of
the simple relation (\ref{r1}).

\section{Estimates on Fractional Moments}\label{fm}\setcounter{equation}{0}

The Aizenman-Molchanov approach of localization for self-adjoint operators
consists in deriving exponential 
estimates on the expectation of fractional powers of the
matrix elements of the resolvent that are uniform in the spectral parameter
\cite{am}. We conduct a similar analysis on the matrix elements of $F(z)$ to prove
Theorem \ref{efm}, following the original strategy and \cite{ag}. 

We restore the dependence in the disorder $\omega$ in the notation at this point and 
we derive the equation satisfied by  the matrix elements $F_\omega(k,j;z)$, $z\not\in S^1$.
We have
\be
\un = (U_\omega-z)(U_\omega-z)^{-1}=(U_\omega-z)S^*F_\omega(z)=(D_\omega -zS^*)F_\omega(z).
\ee
Taking matrix elements, this yields
\bea
\delta_{jk}&=&\bra k |(D_\omega -zS^*)F_\omega(z)j\ket=
e^{-i\theta_k(\omega)}\bra k| F_\omega(z)j\ket -z \bra Sk|F_\omega(z)j\ket\nonumber\\
&=&e^{-i\theta_k(\omega)}F_\omega(k,j;z) -z \sum_{l\in \Z^d}\bra Sk|l\ket F_\omega(l,j;z).
\eea
The diagonal elements of $S=S(t)$ are constant and given by
\be\label{rho}
 \bra Sk|k\ket=(1-t_1^2)(1-t_2^2)\cdots (1-t_d^2)=r_1^2r_2^2\cdots r_d^2\equiv \rho_d(t).
\ee 
Separating the index $l=k$ from the other $l$'s we get for all  $j\neq k$ and 
 $0\neq z\not\in S^1$
\be\label{eqef}
F_\omega(k,j;z)\left( e^{-i\theta_k(\omega)}z^{-1}- \rho_d(t)\right)=
\sum_{l\neq k}\bra Sk|l\ket F_\omega(l,j;z).
\ee

Note that the off-diagonal elements $k\neq l$ satisfy 
\be\label{off}
 \bra Sk|l\ket=\Pi_{j=1}^d\bra S_j(t_j)k_j|l_j\ket=O(|t|), \ \ \ \mbox{ with }\ \ \ 
|t|=\max(t_1,\cdots,t_d),
\ee
since for one $j$ at least, $k_j\neq l_j$, so that there is at least a factor $t_j$ 
in the product, whereas 
\be
 \bra Sk|k\ket=1+O(|t|^2)<1.
\ee 

At this point, we mimick \cite{am} and \cite{ag}. We take $s\in ]0,1[$ and 
try to get estimates on the expectation of $|F_\omega(k,j;z)|^s$. 
Using $|\sum_j a_j|^s\leq \sum_j |a_j|^s$, we infer from (\ref{eqef})
\be
|F_\omega(k,j;z)|^s\left| e^{-i\theta_k(\omega)}z^{-1}-\rho_d(t)\right|^s\leq
\sum_{l\neq k}|\bra Sk|l\ket|^s |F_\omega(l,j;z)|^s, \ \ \ j\neq k.
\ee
Taking expectation and making use of the identity (\ref{b6}) (with $k$ in
place of $j$), this yield
\be\label{rhs}
\E\left(\sum_{l\neq k}|\bra Sk|l\ket|^s |F_\omega(l,j;z)|^s\right)\geq
\E\left( \frac{|\hat F_\omega(k,j;z)|^s\left|e^{-i\theta_k(\omega)}z^{-1}-\rho_d(t)\right|^s}
{|1-\eta_k\hat F_\omega(k,k;z)|^s}
 \right).
\ee
In order to get estimates uniform in $z$, we need to get rid of
the factor $\left|e^{-i\theta_k(\omega)}z^{-1}-\rho_d(t)\right|^s$.
This is done by means of a decoupling lemma similar to the one proven
in \cite{am} for the self-adjoint setting.
Recall that $d\nu(\theta)$ defined on $\T$ is the common distribution of 
the i.i.d. phases $\{\theta_k(\omega)\}_{k\in\Z^d}$.  As $\hat F_\omega$ is independent
of $\theta_k(\omega)$, we shall first average over $\theta_k(\omega)$
and make use of a unitary version of the decoupling Lemma.
\begin{lem}[Decoupling Lemma]
Assume $d\nu(\theta)=\tau(\theta)d\theta$,
where $\tau\in L^\infty(\T)$ is such that $\int_\T d\nu(\theta)=1$.
Then, for any $ 0<s<1$, there exists a constant $0<C_\nu^{(1)}(s)<\infty$ such 
that for all $\alpha, \beta \in\C$
\be\label{1}
\int_{\T}d\nu(\theta)\left|\frac{e^{\pm i\theta} - \alpha }
{e^{\pm i\theta} - \beta}\right|^s\geq C_\nu^{(1)}(s)
\int_{\T}d\nu(\theta)\frac{1 }{\left|e^{\pm i\theta} - \beta\right|^s}.
\ee
Moreover, there exists $0<C_\nu^{(2)}(s)<\infty$ such that for all $\beta\in\C$
\be\label{2}
\int_{\T}d\nu(\theta)\frac{1 } {\left|e^{\pm i\theta} - \beta\right|^s}
\leq C_\nu^{(2)}(s).
\ee
\end{lem}
{\bf Remarks:} \\
i) A proof is provided in Appendix. We only note here that once the estimates hold for 
$e^{i\theta}$ in the integrand, they hold for $ e^{-i\theta}$ by conjugation.\\
ii) A variant of the above result holds for more general 
distributions $d\nu(\theta)$ of phases, in the spirit of \cite{am}, and 
\cite{ag}. \\
iii) As a first application we get the uniform bound
\bea\label{bjj}
\E(|F_\omega(k,k;z)|^s)&=&\E\left(\left|\frac{\hat F_\omega(k,k;z)}
{1-\hat F_\omega(k,k;z) + e^{-i\theta_k(\omega)}\hat F_\omega(k,k;z)}
\right|^s\right)\\
&=&\E\left(\frac{1}{\left|(1-\hat F_\omega(k,k;z))\hat F_\omega(k,k;z)^{-1} + e^{-i\theta_k(\omega)}\right|^s}
\right)\leq C_\nu^{(2)}(s).\nonumber
\eea

We apply now the decoupling Lemma to the RHS of (\ref{rhs}) as follows. We can write
\bea
& &\frac{|\hat F_\omega(k,j;z)|^s\left|e^{-i\theta_k(\omega)}z^{-1}-\rho_d(t)\right|^s}
{|1-\hat F_\omega(k,k;z)+e^{-i\theta_k(\omega)}\hat F_\omega(k,k;z)|^s}=\\
& &\quad \quad \quad \quad \quad \quad \quad  \frac{\rho_d^s(t)|\hat F_\omega(k,j;z)|^s
\left|\rho_d(t)^{-1}z^{-1}-e^{i\theta_k(\omega)}\right|^s}
{|1-\hat F_\omega(k,k;z)|^s\left|e^{i\theta_k(\omega)}+\hat F_\omega(k,k;z)\nonumber
(1-\hat F_\omega(k,k;z))^{-1}\right|^s}.
\eea
Therefore, the average over $\theta_k(\omega)$ of the above yields the bound
\bea
& &\int_\T d\nu(\theta)\frac{|\hat F_\omega(k,j;z)|^s\left|e^{-i\theta_k(\omega)}z^{-1}-\rho_d(t)\right|^s}
{|1-\hat F_\omega(k,k;z)+e^{-i\theta_k(\omega)}\hat F_\omega(k,k;z)|^s}\geq\\
& &\quad \quad \quad \quad \quad  \quad \quad 
C_\nu^{(1)}(s)\rho_d^s(t)\int_\T d\nu(\theta)\frac{|\hat F_\omega(k,j;z)|^s}
{|1-\hat F_\omega(k,k;z)+e^{-i\theta_k(\omega)}\hat F_\omega(k,k;z)|^s},\nonumber
\eea
where the last integrand coincides with $| F(k,j;z)|^s$. Therefore, inserting this in
(\ref{rhs}), we finally get for $j\neq k$ and any $s\in ]0,1[$, 
\be
\sum_{l\neq k}|\bra Sk|l\ket|^s \E(|F_\omega(l,j;z)|^s)\geq C_\nu^{(1)}(s)\rho_d^s(t)
\E(|F_\omega(k,j;z)|^s).
\ee
This last formula is the key to the desired bound, due to the following Lemma,
see \cite{am},\cite{ag}. The proof of \cite{ag} is repeated in Appendix, for completeness.  
\begin{lem}\label{lem}
Let $f\in l^\infty(\Z^d)$ be non-negative and $\sigma : l^\infty(\Z^d)\ra l^\infty(\Z^d)$ 
be a linear operator with kernel $\sigma(k,l)\geq 0$ such that $\sigma(k,k)=0$ and
\be
 \sup_k \sum_{l\neq k}\sigma(k,l)=N <\infty.
\ee
Fix a $j\in\Z^d$ and assume there exists some finite $C>0$ such that $f$ satisfies 
for any $k\neq j$
\be
(\sigma f)(k)=\sum_{l\neq k}\sigma(k,l)f(l)\geq C f(k).
\ee
Then, if $N<C$, and if there exists $\gamma >0$ such that 
\be\label{gam}
\sup_k \sum_{l\neq k}\sigma(k,l)
e^{\gamma|k-l|}<C,
\ee we have for any $k$,
\be
f(k)\leq f(j)e^{-\gamma|j-k|}.
\ee
\end{lem}
This proposition applies to $f(k)=\E(|F_\omega(k,j;z)|^s)$ and
$\sigma(j,k) = |\bra S j|k\ket|^s$ with the constants
\be\label{cn}
C=C_\nu^{(1)}(s)\rho_d^s(t), \ \ \ \mbox{ and } \ \ \ 
N=\sup_k \sum_{j\neq k} |\bra S k|j\ket|^s,
\ee
for small enough values of $|t|$.
Indeed, for $z\not\in S^1$, we have the {\it a priori} bound 
\be
|F_\omega(k,j;z)|=|\bra j |S(U_\omega -z)^{-1} k\ket|\leq 1/\mbox{ dist}(z, S^1),
\ee
showing that $f$ is in $l^\infty$. Moreover, if $|t|$ is small enough,
we get from (\ref{rho}) and (\ref{off}) that 
\be
N=\sup_k \sum_{j\neq k} |\bra S k|j\ket|^s =O(|t|^s)< C_\nu^{(1)}(s)\rho_d^s(t)=
C_\nu^{(1)}(s)+O(|t|^2)=C.
\ee
Finally, as the sum defining $N$ in (\ref{cn}) carries 
over a finite number of indices only, for such values of $|t|$, there exists
a $\gamma=\gamma(s,t)>0$ so that (\ref{gam}) holds true. 
With the uniform bound 
on $\E(|F_\omega(j,j;z)|^s)$ derived in (\ref{bjj}), and by the fact that $D_\omega$ is
diagonal, this ends the proof of Theorem \ref{efm}.\hfill \ep

\section{Localization}\label{loc}\setcounter{equation}{0}

We spell out here a spectral consequence of the estimates derived in 
Theorem \ref{efm} by proving  Corollary \ref{uloc}. 
We do this by applying the unitary version of the 
Simon-Wolff criterion \cite{sw} for localization presented by Combescure in 
\cite{c}, see also \cite{t}. 

\vspace{.3cm}

We need some preliminary estimates.
Let us introduce for $z\not\in S^1$,
\be
H_\omega(z)=U_\omega (U_\omega -z)^{-1}.
\ee
We choose $j=0$ in the definition (\ref{uhat}).
By the Spectral Theorem ,
\be
H_\omega (z)=\int_\T \frac{dE_\omega (\alpha)e^{i\alpha}}{e^{i\alpha}-z},
\ee
where $E_\omega (\alpha)$ is the spectral family associated with $U_\omega $. Therefore,
the spectral measure associated with $|0\ket$
\be
d\mu_\omega(\alpha)= d\bra 0|E_\omega (\alpha) 0\ket =d \| E_\omega (\alpha) 0\ket\|^2
\ee
is such that
\be
\bra 0|H_\omega (z) 0\ket = \int_\T \frac{d\mu_\omega(\alpha)e^{i\alpha}}{e^{i\alpha}-z}.
\ee
Thus, for $z<1$,
\be
\|H_\omega (z) 0\ket \|^2=\bra 0| H_\omega ^*(\bar z) H_\omega (z) 0\ket=  
\int_\T \frac{d\mu_\omega(\alpha)}{|e^{i\alpha}-z|^2}.
\ee
Introducing the Poisson integral of a measure $d\mu$ 
\be
P[d\mu](z)=\int_\T \frac{d\mu(\alpha)(1-|z|^2)}{|e^{i\alpha}-z|^2}\geq 0, \ \ \ |z|< 1, 
\ee
the identity above for $z=re^{i\theta}, r<1$ can be cast under the form 
\bea
\|H_\omega (re^{i\theta}) 0\ket \|^2&=&P[d\mu_\omega](re^{i\theta})+
\int_\T \frac{d\mu_\omega(\alpha)r^2}{1+r^2-2r\cos(\alpha -\theta)}\\
&\equiv&P[d\mu_\omega](re^{i\theta})+B_\omega(r,\theta).\label{br}
\eea
We know that the following limit exists and is finite for a.e. 
$\theta\in\T$ with respect to $d\theta/2\pi$ 
\be
\lim_{r\ra 1^-}P[d\mu_\omega](re^{i\theta})=\frac{d\mu_\omega(\theta)}{d\theta}.
\ee
Since
\be\label{mono}
r\mapsto \frac{r^2}{1+r^2-2r\cos(\alpha -\theta)} \ \ \ 
\mbox{is positive, monotone increasing,} 
\ee
then
\be\label{bt}
\lim_{r\ra 1^-}B_\omega(r,\theta)=
\int_\T \frac{d\mu_\omega(\alpha)}{4\sin^2((\alpha-\theta)/2)}\equiv B_\omega(\theta)
\ \ \ \mbox{exists for all $(\omega,\theta)\in \Omega\times \T$.} 
\ee
Moreover,  for $(\omega, \theta)$ fixed, $B_\omega(r,\theta)$
is monotone non-decreasing in $r$ as well.
Now, Theorem \ref{efm} says for $0<s<1$, 
\be
\E(|\bra j |H_\omega (z)0\ket|^s)\leq K(s) e^{-\gamma(s)|j|}, \ \ \ 
\mbox{uniformly in $z$}.
\ee
Together with
\be
\left(\sum_j|\bra j |H_\omega (z)0\ket|^2\right)^{\tilde s}\leq \sum_j
|\bra j |H_\omega (z)0\ket|^{s},  \ \ \ \mbox{for $\tilde s = s/2<1$,}
\ee
this implies 
\be
\E\left(\left(\|H_\omega (z)0\ket\|^{2}\right)^{\tilde s}\right)\leq
\sum_j K(s) e^{-\gamma(s)|j|}=\tilde K(s)<\infty .
\ee
Thus, we can apply the Monotone Convergence Theorem again to (\ref{bt})
for the measure $d\theta \times d\P(\omega)$ to get from (\ref{br}) 
that,
\bea
\int_\T d\theta \ \E( (B_\omega(\theta))^{\tilde s})&=&\lim_{r\ra 1^-}
\int_\T d\theta \ \E( (B_\omega(r,\theta))^{\tilde s})\\
&\leq& \lim_{r\ra 1^-}\int_\T d\theta \ \E\left(\left( B_\omega(r,\theta)+
P[d\mu_\omega](re^{i\theta}) 
\right)^{\tilde s}\right)\nonumber\\
&=& \lim_{r\ra 1^-}\int_\T d\theta \ \E\left(\left( \|H_\omega (re^{i\theta})0
\ket\|^{2}\right)^{\tilde s}\right)\leq 2\pi\tilde K(s).
\eea
Therefore,  $ B_\omega(\theta)$ is finite for almost all $(\theta, \omega) 
\in \T\times \Omega $, w.r.t. $d\theta \times d\P(\omega)$. By Fubini, 
this implies 
\begin{prop}\label{boundb} Under the hypotheses of Theorem \ref{efm}, 
there exists $\tilde \Omega_0\subset \Omega$ of probability one and
$J_\omega\in \T$ of full measure such that 
\be
B_\omega(\theta)<\infty \ \ \ \mbox{ if } \ \ \omega \in \tilde \Omega_0 \ \ 
\mbox{ and }\ \  \theta \in J_\omega.
\ee
\end{prop}

We are now in a position to apply the unitary version of \cite{c} of the
Simon-Wolff criterion for localization. Consider
\be
\hat H_\omega (z)=\hat U_\omega (\hat U_\omega  -z)^{-1} \ \ 
\ \mbox{corresponding to (\ref{uhat})}
\ee
and $d\hat \mu_\omega$ the corresponding spectral measure associated with the vector
$|0\ket$. The relation (\ref{b6}) for $j=k=0$ is equivalent to
\be\label{relhat}
\bra 0 |H_\omega(z) 0\ket=\frac{\bra 0 |\hat H_\omega(z) 0\ket}{\bra 0 |\hat H_\omega(z) 0\ket
(1-e^{i\theta_0(\omega)})+e^{i\theta_0(\omega)}}.
\ee
The properties of the perturbed spectral measure $d \mu_\omega$, i.e. with $\theta_0(\omega)$
arbitrary,  can be read from 
those of the unperturbed  spectral measure $d\hat \mu_\omega$, i.e. with $\theta_0(\omega)=0$,
by means of the unitary analog of the Aronszajn-Donoghue 
characterization of supports of the Lebesgue decomposition of the
spectral measure $d\mu_\omega$. We recall this characterization 
for completeness, 
changing slightly notations with respect to \cite{c}: Combescure uses
the resolvent rather  than $H_\omega(z)=1+z(U_\omega -z)^{-1}$. 
Let $\hat B_\omega(\theta)$ be defined by (\ref{bt}) 
for $\hat d\mu_\omega$ in place of $d\mu_\omega$.
\begin{prop}
With the notations above,\\ 
a support of the singular continuous part of $d\mu_\omega$ is
\be
S_\omega =\left\{\theta\in\T \ | \lim_{r\ra 1^-}
\hat H_\omega(re^{i\theta})=\frac{e^{i\theta_0}}{e^{i\theta_0}-1} 
\mbox{ and }  \hat B_\omega(\theta)=\infty \right\},
\ee
 the set of atoms of $d\mu_\omega$ is
\be
P_\omega =\left\{\theta\in\T \ | \lim_{r\ra 1^-}
\hat H_\omega(re^{i\theta})=\frac{e^{i\theta_0}}{e^{i\theta_0}-1} 
\mbox{ and }  \hat B_\omega(\theta)<\infty \right\},
\ee
whereas a support of the absolutely continuous part of $d\mu_\omega$ is
\be
A_\omega =\left\{\theta\in\T \ | \lim_{r\ra 1^-}P[d\mu_\omega](re^{i\theta})=
\frac{d\mu_\omega(\theta) }{d\theta}\in (0,\infty) \right\}.
\ee
These sets are mutually disjoint. 
\end{prop}
The key proposition from \cite{c} regarding the properties of $d \mu_\omega$
in our setting  is the following  unitary version of the Simon-Wolff criterion:
\begin{prop} Let $d \hat \mu_\omega$ and $d\mu_\omega$ be related by 
(\ref{relhat}).
\be\hat B_\omega(\theta)< \infty \ \, \mbox{for a.e $\theta\in\T$} \Longleftrightarrow 
d\mu_\omega \ \, \mbox{is purely atomic for a.e. $\theta_0 \in\T$}. 
\ee
\end{prop}

Indeed, considering $\hat U_\omega:=e^{i\theta_0(\omega)}U_\omega$ instead of 
$U_\omega$, we deduce from  Proposition \ref{boundb} and the criterion above 
that for any $\omega\in \tilde \Omega_0$, 
the spectral measure for $|0\ket $ of
\be
\tilde U_\omega= e^{-i\beta|0\ket\bra 0|}U_\omega=\mbox{diag} 
(e^{-i \theta_{j}^\beta(\omega)})S, \ \ \ \mbox{ where } \ \ \
\theta_{j}^\beta(\omega)=\theta_{j}(\omega)+\beta \delta_{j,0},
\ee
is purely atomic for almost all $\beta\in\T$. But, as the distribution of
phases is absolutely continuous, this means that the spectral measure 
$d\mu_\omega(\cdot)= \bra 0 | dE(\cdot) 0\ket$ of $U_\omega$ is purely atomic
for $\omega\in \Omega_0$, a set of probability one. Repeating the 
argument for the spectral measures
$\bra j | E(\cdot) j \ket $, $j\in\Z^d$, this yields the same result 
for $\omega\in\Omega_j$, where $\Omega_j$ is a set of probability one. 
Therefore, $U_\omega$ is pure point for $\omega\in\cap_{j\in\Z^d}\Omega_j$, a
set of probability one.\\ \phantom{p}\hfill \ep

\section{Appendix}\setcounter{equation}{0}

\subsection{Proof of the Decoupling Lemma}
Let us start with the second part of the Lemma. For any  $\lambda >0$,
\bea\label{ub}
\int_\T \frac{d\nu(\theta)}{|e^{i\theta}-\beta|^s}&\leq& 
\lambda \int_{\{|e^{i\theta}-\beta|^{-s}\leq \lambda\}}d\nu(\theta)+
\int_{\{|e^{i\theta}-\beta|^{-s}\geq \lambda\}}\frac{d\nu(\theta)}{|e^{i\theta}-\beta|^{s}}\nonumber\\
&\leq&\lambda + \int_{\lambda}^\infty \nu\{|e^{i\theta}-\beta|^{-s}\geq \lambda'\}d\lambda',
\eea
where 
\be
\nu\{|e^{i\theta}-\beta|^{-s}\geq \lambda\}\leq \|\tau\|_\infty\int_{\{|e^{i\theta}-\beta|\leq 
1/\lambda^{1/s}\}}d\theta.
\ee
In the last integral, we can assume without loss that $\beta\geq 0$ and it 
yields the arclength of the intersection of 
the unit circle with a circle of radius 
$1/\lambda^{1/s}$, centered at $\beta$. We first note that 
if $\lambda^{-1/s}\geq 1$, i.e. $\lambda\leq 1$, the integral takes its maximal value $2\pi$, 
obtained with $\beta=0$. If $\lambda >1$, the integral is maximized by the choice
$\beta=\beta(\lambda)=\sqrt{1-\lambda^{-2/s}}$ to give 
\be\label{cola}
\int_{\{|e^{i\theta}-\beta|\leq 1/\lambda^{1/s}\}}d\theta =
2\arcsin(1/\lambda^{1/s}), \ \ \ \mbox{ if }\ \ \ 1/\lambda^{1/s}< 1.
\ee
As $\lambda\ra \infty$, this integral behaves as $1/\lambda^{1/s}$, which 
is integrable for $0<s<1$. At this point we optimize our upper bound 
(\ref{ub}) on $\lambda$ by choosing $\lambda$ such that
\be
1-\|\tau\|_\infty 2\arcsin(1/\lambda^{1/s})=0.
\ee
Since $\|\tau\|_\infty\geq 2\pi$ to ensure normalization, the minimizer
is 
\be
\lambda=[\sin(1/(2\|\tau\|_\infty)))]^{-s}>1.
\ee
Therefore we have proven the existence of a constant 
$C_\nu^{(2)}(s)$ depending on $s$ and $\nu$ only such that (\ref{2}) holds.

The first part of the Lemma is  proven along the lines of  \cite{ag}. It is shown  
in the appendix C of this paper that for $0<s<1$ and for any 
$u,v,\alpha,\beta\in \C$, 
\bea
\frac{1}{|v-\beta|^s}+\frac{1}{|u-\beta|^s}&\leq& \frac{|v-\alpha|^s}{|v-\beta|^s}
\left(\frac{1}{|u-\alpha|^s}+\frac{1}{|u-\beta|^s}\right)\nonumber\\
&+& \frac{|u-\alpha|^s}{|u-\beta|^s}
\left(\frac{1}{|v-\alpha|^s}+\frac{1}{|v-\beta|^s}\right).
\eea
Then, replacing $v$ and $u$ by $e^{i\theta}$ and  $e^{i\theta'}$ respectively,
and integrating over $d\nu(\theta)d\nu(\theta')$, we get
\bea
& &\int_\T\int_\T d\nu(\theta)d\nu(\theta')\left(
\frac{1}{|e^{i\theta}-\beta|^s}+\frac{1}{|e^{i\theta'}-\beta|^s}\right)=
2\int_\T d\nu(\theta)\frac{1}{|e^{i\theta}-\beta|^s}\nonumber\\
& \leq&\int_\T d\nu(\theta)\frac{|e^{i\theta}-\alpha|^s}{|e^{i\theta}-\beta|^s}
\int_\T d\nu(\theta')\left(\frac{1}{|e^{i\theta'}-\alpha|^s}+\frac{1}{|e^{i\theta'}-\beta|^s}\right)
+ (\theta' \leftrightarrow \theta )
\eea
where $(\theta' \leftrightarrow \theta )$ means the same expression with $\theta$ and 
$\theta'$ exchanged. We finally get (\ref{1}) with $C^{(1)}_\nu(s)=1/(2C^{(2)}_\nu(s))$ by applying the bound (\ref{2}).\ep 
\subsection{Proof of Lemma \ref{lem}}
We first observe that if $\ffi =\{\ffi(k)\}_{k\in\Z^d}\in L^\infty(\Z^d)$ is real valued,
such that $\ffi(j)\leq 0$ and satisfies 
\be
C\ffi(k)\leq (\sigma \ffi)(k), \ \ \ \forall k\neq j,
\ee
then $\ffi(k)\leq 0$, for any $k$. Indeed, if it were not the case, 
$M\equiv \sup_{k}\ffi(k)$ would be strictly positive. But that would imply
\be
C\ffi(k)\leq \sum_{l\neq k}\sigma(k,l)\ffi(l)\leq N M \ \Rightarrow C M\leq N M,
\ee
which contradicts $N<C$. Then one applies the above to
\be
\ffi(k)=f(k)-f(j)e^{-\gamma|k-j|}, \ \ \ \mbox{s. t. } \ \ \ffi(j)=0.
\ee
Since 
\bea
(\sigma e^{-\gamma|\cdot -j|})(k)&=&\sum_{l\neq k}\sigma(k,l)e^{-\gamma(|l-j|-|k-j|)}
e^{-\gamma|k-j|}\leq \sum_{l\neq k}\sigma(k,l)e^{\gamma|l-k|}e^{-\gamma|k-j|}\nonumber\\
&\leq& Ce^{-\gamma|k-j|},
\eea
by hypothesis, we get, using $f(j)\leq 0$,
\be
(\sigma\ffi)(k) = (\sigma f)(k)-f(j)(\sigma e^{-\gamma|\cdot -j|})(k)\geq 
C(f(k)-f(j)e^{-\gamma|k-j|})=C \ffi(k),
\ee
hence $f(k)\leq f(j)e^{-\gamma|k-j|}$. \hfill\ep


\vspace{.3cm}

 \begin{thebibliography}{xxxxxxx}
\bibitem[AM]{am} Aizenman, M., Molchanov, S. : Localization at  large disorder
and at extreme energies: an elementary derivation., {\it Commun. Math. Phys.}
{\bf 157}, 245-278, (1993).
\bibitem[AG]{ag} Aizenman, M., Graf G.-M. : Localization Bounds for an
Electron Gas, {\it J. Phys. A} {\bf 31}, 6783-6806, 
(1998).
\bibitem[BB]{bb} G. Blatter, D. Browne, {\it Zener tunneling and
    localization in small conducting rings}, Phys. Rev. B, {\bf 37}, 
(1988), 3856.
\bibitem[BHJ]{bhj} Bourget, O., Howland, J.S., Joye, A. :  "Spectral Analysis of Unitary 
Band Matrices", {\it Commun. Math. Phys.}, {\bf 234} , (2003), p. 191-227 . 
\bibitem[C]{c} Combescure, M. : Spectral Properties of a Periodically Kicked 
Quantum Hamiltonian, {\it J. Stat. Phys.}{\bf 59}, 679-690, (1990)
\bibitem[CMV]{cmv} M.J. Cantero, L. Moral and L. Vel\'azquez,
{\it Five-Diagonal Matrices and Zeros of Orthogonal Polynomials
on the Unit Circle}, Linear Algebra and Its Applications, {\bf 326 }C, 29-56, (2003) 
\bibitem[GT]{gt} J.S. Geronimo, A. Teplyaev, {\it A Difference Equation
Arising from the Trigonometric Moment Problem Having Random Reflection 
Coefficients-An Operator Theoretic Approach}, J. Func. Anal., {\bf 123},
(1994), 12-45. 
\bibitem[J]{j} Joye, A. : Density of States and Thouless Formula
for Random Unitary Band Matrices, {\it Ann. Henri Poincar\'e} {\bf 5},
347--379, (2004).
\bibitem[S1]{s} Simon, B.: Orthogonal Polynomials on the Unit Circle, 
Vol. 1 and 2, {\it AMS Colloquium Series, American Mathematical 
Society, Providence, RI}, to appear.
\bibitem[S2]{ps} Simon, B.: Aizenman's Theorem for Orthogonal
Polynomials on the Unit Circle, preprint, mp-arc 04-386.
\bibitem[SW]{sw} Simon, B., Wolff, T.: Singular Continuous Spectrum under 
Rank One Perturbations and Localization for Random Hamiltonians, {\it Commun. 
Pure Appl. Math.} {\bf 39}, 75--90, (1986).
\bibitem[T]{t} Teplyaev, A. V., the Pure Point Spectrum of Random Polynomials
orthogonal on the Circle, {\it Soviet. Math. Dokl.}, {\bf 44}, 407-411, (1992).

 \end{thebibliography}
\end{document}